\documentclass[journal]{IAENGtran}
\usepackage{cite}

\ifCLASSINFOpdf
   \usepackage[pdftex]{graphicx}
   \DeclareGraphicsExtensions{.pdf,.jpeg,.png}
\else
   \usepackage[dvips]{graphicx}
   \DeclareGraphicsExtensions{.eps}
\fi

\usepackage[cmex10]{amsmath}
\usepackage{amssymb}

\newcommand\argmax{\ensuremath{ \mathop{ \arg\!\max  } } }
\usepackage{color}

\begin{document}

\title{Optimal Data Split Methodology for Model Validation}

%

\author{Rebecca Morrison, Corey Bryant, Gabriel Terejanu, Kenji Miki, Serge Prudhomme

\thanks{Manuscript received June 30, 2011; revised August 13, 2011. This material is based upon work supported by the
Department of Energy [National Nuclear Security Administration] under Award Number [DE-FC52-08NA28615].}

\thanks{All authors are located at the Institute for Computational Engineering and Sciences, The University of Texas at
Austin; Austin, TX, 78712 e-mail: \{rebeccam, cbryant, terejanu, kenji, serge\}@ices.utexas.edu.}
} 

%
%

%

\maketitle

\pagestyle{empty}
\thispagestyle{empty}

\begin{abstract}
The decision to incorporate cross-validation into validation processes of mathematical models raises an immediate
question -- how should one partition the data into calibration and validation sets? We answer this question
systematically: we present an algorithm to find the optimal partition of the data subject to certain constraints. While
doing this, we address two critical issues: 1) that the model be evaluated with respect to predictions of a given
quantity of interest and its ability to reproduce the data, and 2) that the model be highly challenged by the validation
set, assuming it is properly informed by the calibration set. This framework also relies on the interaction between the
experimentalist and/or modeler, who understand the physical system and the limitations of the model; the decision-maker,
who understands and can quantify the cost of model failure; and the computational scientists, who strive to determine if
the model satisfies both the modeler's and decision-maker's requirements. We also note that our framework is quite
general, and may be applied to a wide range of problems. Here, we illustrate it through a specific example involving a
data reduction model for an ICCD camera from a shock-tube experiment located at the NASA Ames Research Center (ARC).

\end{abstract}

\begin{IAENGkeywords}
 Model validation, quantity of interest, Bayesian inference
\end{IAENGkeywords}

\IAENGpeerreviewmaketitle


\section{Introduction} Model validation to assess the credibility of a given model is becoming a necessary activity when
making critical decisions based on the results of computer modeling and simulations. As stated in \cite{challenge},
validation requires that the model accurately capture the critical behavior of the system(s), and that uncertainties due
to random effects be quantified and correctly propagated through the model.

There are various approaches to model validation; here we explore a procedure based on cross-validation. First, one
partitions the data into two sets: the \textit{calibration} (or training) set and the \textit{validation} set. Next, the
calibration set is used to calibrate the model. Then the calibrated model produces a set of predicted values
to be compared with the validation set. A small discrepancy between predicted values and the validation set improves the
credibility of the model while a large discrepancy may invalidate the model. We will present a general framework based
on these principles that incorporates our particular goals along with a detailed cross-validation algorithm.

More specifically, we examine situations in which we want to predict values for which experimental data is not
available, referred to here as the \textit{prediction scenario}. Experiments for this scenario may be impractical or even
impossible. Still, as a computational scientist, one wants to predict a certain quantity of interest (QoI) at this
scenario and assess the quality of this prediction. Often, the only experimental data available comes from legacy
experiments and may be incomplete. Furthermore, this QoI is seldom directly observable from the system, but requires
some additional modeling.

Numerous examples of this situation exist. Computational models aimed at predicting the reentry of space vehicles are
one such example. Some characteristics of the system can be recreated in sophisticated wind tunnels, but experiments are
expensive and may be unreliable.
Another example is the maintenance of nuclear stockpiles. Since experiments to assess environmental impact in case of
failure are banned, predictive models must be used.  

Babu\v{s}ka et al. present a systematic approach to assess predictions of this type using Bayesian inference and what they
call a validation pyramid \cite{Babuska2008}. Scenarios of varying complexity are available that suggest an obvious
hierarchy on which to validate the model.  In their calibration phase, Bayesian updating is used to condition the model
on the observations available at lower levels of the pyramid. The model's predictive ability is then assessed by further
conditioning using the validation data at the higher levels.  One advantage of their approach is that the prediction
metric is directly related to the QoI. This feature is maintained in our work. 

In the work described above, the authors employ a single split of the data into calibration and validation scenarios.
While they argue that this partition of the data is made clear by the experimental set-up and validation pyramid, it is
often the case that all experiments provide an equal amount of information regarding the QoI. To avoid a subjective
choice of the calibration set, we determine the set by a more rigorous and quantitative process.

To do this, we propose a cross-validation inspired methodology to partition the data into calibration and validation
sets. In contrast to previous works, we do not immediately choose a single partition as above, nor do we use averages
or estimators over multiple splits (see \cite{Vehtari2002, Arlot2010, Alqallaf2001} and references therein). Instead,
our approach first considers all possible ways to split the data into disjoint calibration and validation sets which
satisfy a chosen set size. Then, by analyzing several splits, we methodically choose what we term the ``optimal'' split.
Once the optimal split is found, we are then able to judge the validity of the model, and whether or not it should be
used for predictive purposes. 


First, we argue that the model's ability to replicate the observations must be assessed quantitatively. A model
incapable of reproducing observations should not be used to predict unobservable quantities of interest. Specifically in
the case of Bayesian updating, the prior information and observed data must be sufficient to produce a satisfactory
posterior distribution for the model parameters. 

Second, we address a fundamental issue of validation. We can never fully validate a model; instead, we
can only try to invalidate it. Because of this, when we choose a validation set with which to test the model, it should 
be the most challenging possible. In other words, we demand that the model perform well on even the most
challenging of validation sets; otherwise, we cannot be confident in its prediction of the QoI.

With these concepts in mind, we propose that the optimal split satisfy the following desiderata:
\begin{enumerate}
\item[(I)] The model is sufficiently informed by the calibration set (and is thus able to reproduce the data).
\item[(II)] The validation set challenges the model as much as possible with respect to the quantity of interest.  
\end{enumerate}
Using the optimal partition, we are then able to answer whether the model should be used for prediction.

In the current work, we apply our framework to a data reduction model which converts photon counts received by an ICCD camera to
radiative intensities \cite{camera}. We chose this as an appropriate application
because the resulting intensities are later used to make higher-level decisions. Thus, the validity of the model should
be thoroughly tested before the model is deemed reliable. Moreover, the uncertainty in such a model should be explicitly
evaluated, as possible errors may be propagated through to other predicted quantities.

The paper is organized as follows. In section II, the general framework is detailed step by step. A concise algorithm is
provided. In section III, the approach is applied to the data reduction model. Finally, in section IV, short-comings and
future work are discussed. 

\section{Validation Framework}

\begin{figure}[h]
 \center
 \includegraphics[width=3.5in]{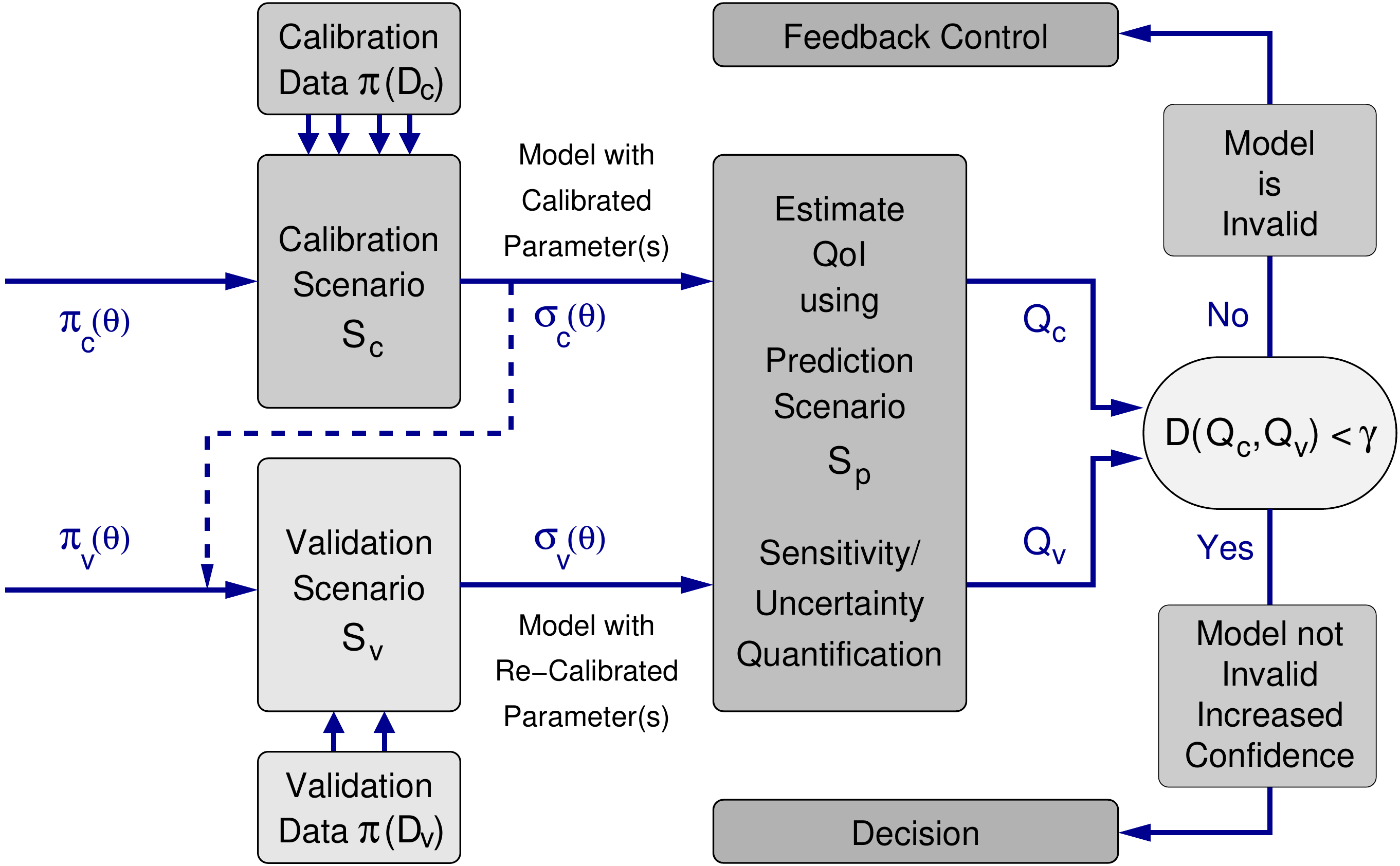}
 \caption{The calibration and validation cycle}
 \label{fig:valCycle}
\end{figure}

Figure \ref{fig:valCycle} demonstrates the previously described framework \cite{Babuska2008}, which applies when the
quantity of interest is only available as a prediction through the computational model, not through direct observation.
There, Bayesian updating is performed on a calibration set, and then a prediction of the QoI is made using the updated
model. A subsequent update is performed using a validation set followed by an additional prediction with the newly
updated model. Finally, the two predictions are compared to assess the model's predictive capabilities.

\subsection{Prediction metric}

The QoI driven model assessment developed in \cite{Babuska2008} requires a metric comparing predictions of the QoI
obtained from the calibration and validation sets. In many instances the QoI is determined by a decision-maker, who may
not be the computational scientist performing the analysis. Consequently, they must work together to develop a suitable
metric for the QoI, as well as a tolerance, which quantifies how consistent the predictions must be (note that we
cannot measure accuracy of the predictions because we have no true value). Ideally this metric would measure in the
units of the QoI, or provide a relative error, allowing for easy interpretation by decision-makers. Examples of such
metrics include absolute error measures and percent error based on a nominal value. 

In the following sections of the paper we will denote the metric used to measure the predictive performance of the model
by $M_Q$, highlighting the fact that it is determined by the QoI. Likewise, we will denote the threshold, or tolerance,
by $M^*_Q$. We stress that generality is maintained since the rest of the procedure discussed below is flexible to the
particular choice of metric and threshold. What we do require is that the choice of metric be appropriate for the
application at hand.  For instance, when using Bayesian inference we require that the metric be compatible with
probabilistic inputs, since predictions are provided as probability or cumulative density functions of the QoI.

\subsection{Data reproduction}
As discussed briefly in the introduction, one must also ensure that the model is capable of reproducing the observed
data. This evaluation has been overlooked, or at least not emphasized, in previous works \cite{Babuska2008, camera}. Such an
evaluation provides confidence that the model and data provided are mutually suitable. Verifying the model's
reproducibility of the observables demonstrates that the parameters of the model are adequately informed through the
inverse problem. 

As in the case of the prediction metric, establishing a performance criterion may require further knowledge of the
physical system. The analyst is likely not an expert in the system being modeled and should solicit a modeler's, or
experimentalist's, assistance in developing a performance metric and tolerance. Doing so will certify that the correct
aspects of the system are being captured sufficiently by the model. 

We use a similar notation as for the prediction metric. Let $M_D$ and $M^*_D$ denote the data reproduction
metric and threshold respectively. We note again that these choices are defined based on the model being considered and
are not specific to the framework we propose. 

\subsection{Choice of calibration size}
As with many validation schemes, the selection of the calibration, or training, set size is important. From $k$-fold to
leave-one-out cross validation, this set size can vary greatly \cite{Arlot2010, Alqallaf2001}. Here, we do not require a
particular choice as the proposed framework will apply whatever this size may be. Indeed, various choices were
considered in preparation of this work; however, we do not discuss them further. 

With that being said, we do recognize that the particular choice could impact the final conclusion and must be made with
care. Of particular concern is providing enough data so that all parameters of the model are sufficiently informed by
the inverse problem. If the model were to fail with respect to the data metric, the issue may not be the model itself
but too small a calibration set size. In this case one should perform further analysis to determine the source of this
discrepancy and increase the calibration set size if necessary. 

Given $N$ observations, we denote the size of the calibration set by $N_C$ and the size of the validation set by $N_V$.
That is, 
\begin{equation}
 N_C + N_V = N.
 \label{eq:calSetSize}
\end{equation}
Note that it could be the case that each of the $N$ observations in fact represents a set of observations, if, for
instance, repeated experimental measurements are taken at the same conditions.

We do not perform our analysis on a single partitioning of the data but
instead consider all possible partitions of the data respecting \eqref{eq:calSetSize}. 
The reason for this approach is two-fold. First, it reduces the sensitivity of the final outcome to any particular set
of data. Since each data point is equivalent under partitioning, we do not bias the groupings in any subjective way
(once we have chosen the calibration set size).

Second, we envision an application where it is unclear which experiments relate more closely to the QoI scenario.
Considering all admissible partitions can provide insight as to which observations are most influential with respect to
the QoI. As an example, consider a case of resonance where the QoI is associated with the resonant behavior of the
system. Without knowing a priori the resonance frequency of the
system, one cannot say which frequencies will be important for capturing the resonant behavior. 

However, the drawback of this approach is evident: we must consider the model performance for all partitions of the data.
This yields a combinatorially large number of partitions, whose exact number is given by the binomial formula:
\begin{equation}
  P = {N \choose N_C} = \frac{N!}{N_C!N_V!}.
  \label{eq:partitions}
\end{equation}
We will denote these partitions, or splits, by $\{s_k\}$, where $k=1,2,\ldots,P$.  The computational impact of this
becomes even more significant while performing the next step of the procedure. 

\subsection{Inversion for model parameters}
For each admissible partition of the data, we solve an inverse problem using the calibration set, of size $N_C$, as input
data. It is not hard to see why solving $P$ inverse problems may be difficult, or even impossible, for
complicated models. This is an area for improvement; approximations and alternative approaches to reduce the number of inverse problems will be the subject of future work. 

As discussed previously, we treat the inverse problem in a probabilistic setting, using Bayesian updating to incorporate
the calibration data. As a result we obtain distributions of model parameters, and these in turn yield distributions for
the predicted quantities. At this point it becomes clear that the definition of the metrics will depend on how the
inverse problems are solved. Note, of course, that a deterministic approach could also be used. 

\subsection{Computation of the metrics}
\begin{figure}[h]
  \begin{center}
    \includegraphics[angle=-90,width=3in]{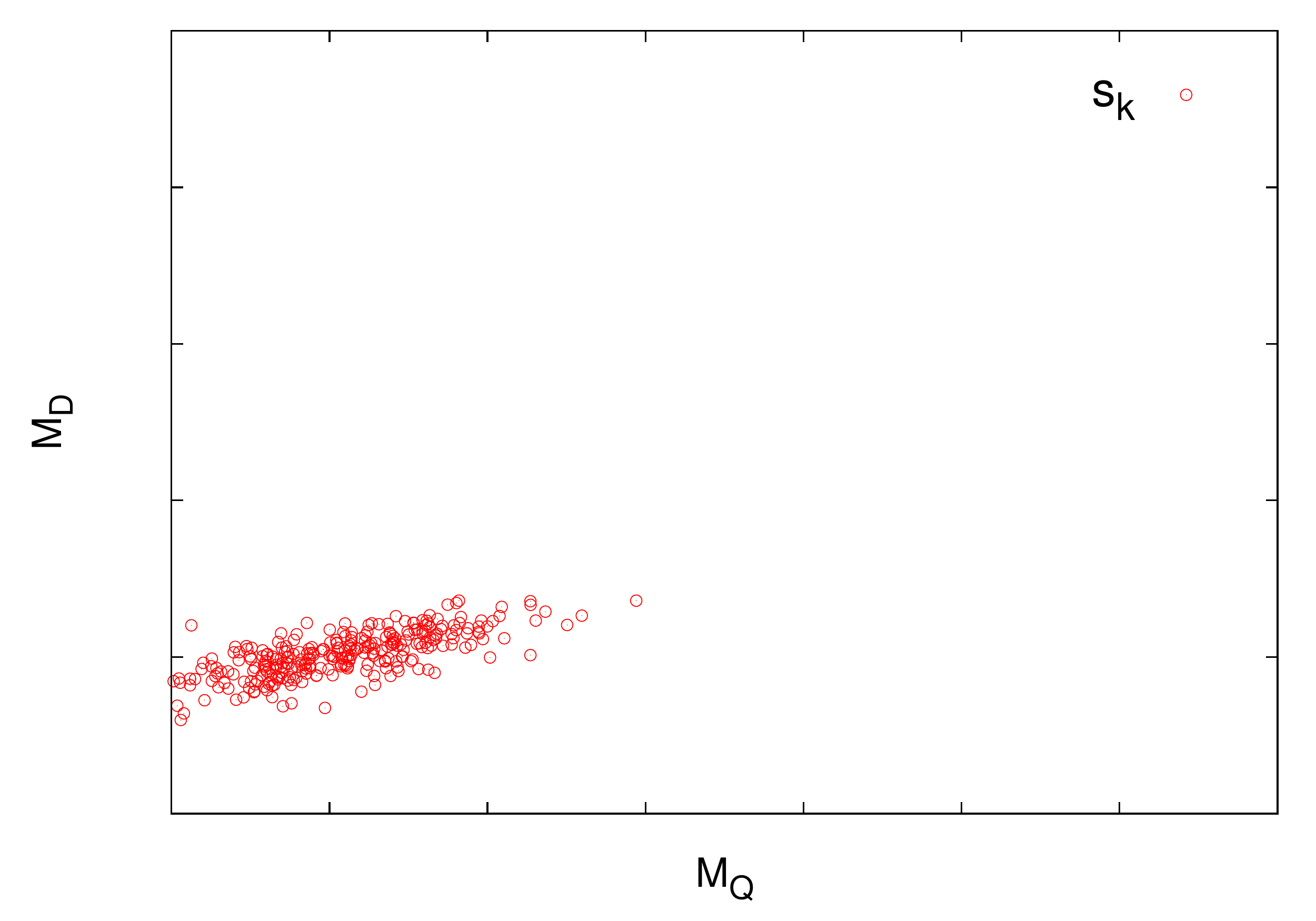}
  \end{center}
  \caption{Metrics computed for each data split $s_k$}
  \label{fig:step6}
\end{figure}

With the solutions obtained from the inverse problems we are now able to evaluate the model's performance. For each calibration set, we compute the metrics as detailed above. One can then
visualize the data on a Cartesian grid where the $x$ and $y$ axes correspond to the metrics $M_Q$ and $M_D$,
respectively, and each point corresponds to a single
partition of the data into a calibration and validation set (figure \ref{fig:step6}).

\subsection{Optimal partition} 
We attempt to invalidate the model using the optimal split determined by (I) and (II).

For (I), performance in replicating the observables is measured using the data metric $M_D$. Thus we only consider splits, $s_k$, of the data that satisfy
\begin{equation*}
  M_D(s_k) < M^*_D.
\end{equation*}
If no points lie below this threshold, the model's ability of reproducing the observed data is unsatisfactory, and one
must change or improve the model, or the data, or perhaps both. 

\begin{figure}[h]
  \begin{center}
    \includegraphics[angle=-90,width=3in]{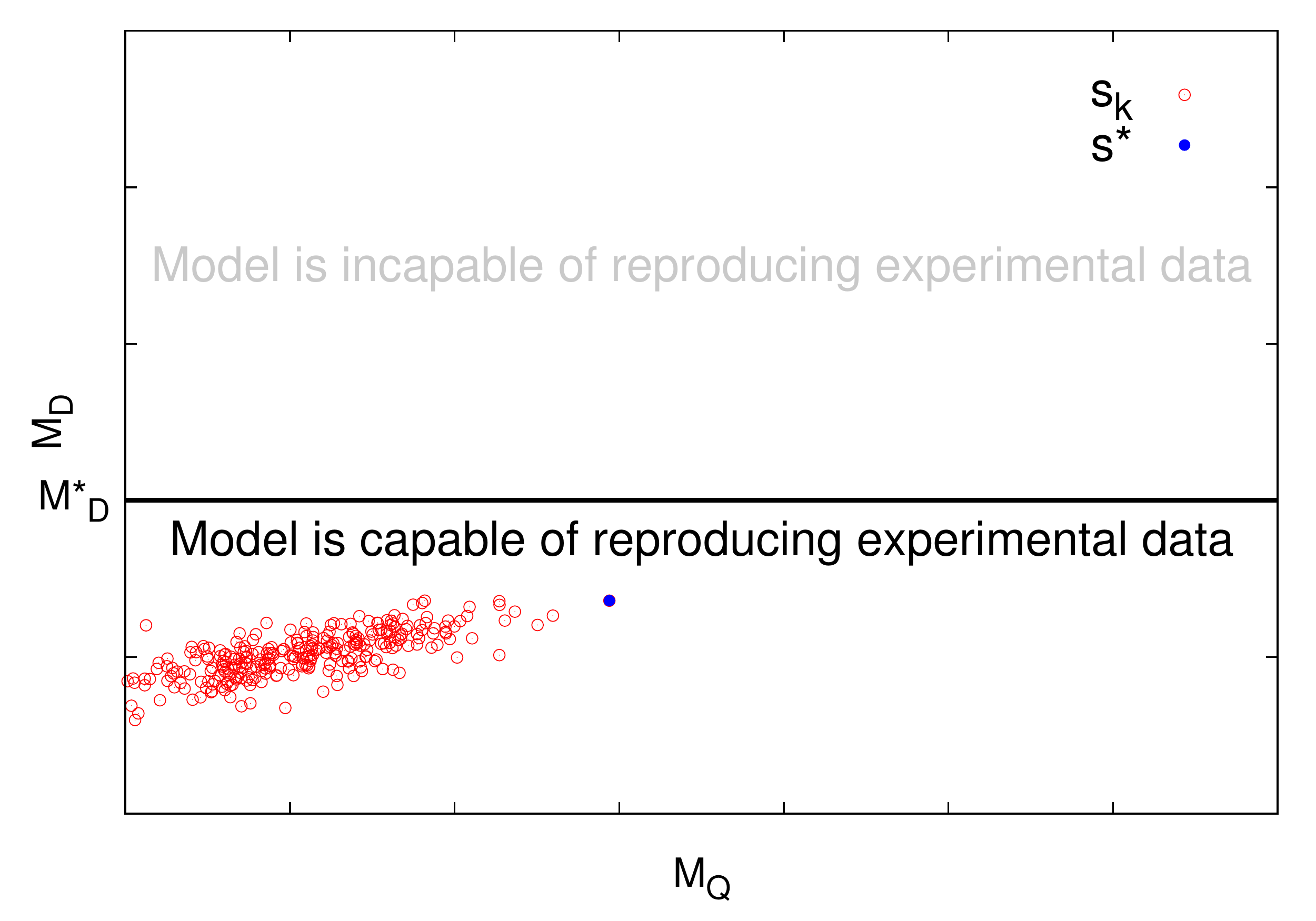}
  \end{center}
  \caption{Identification of optimal split $s^*$}
  \label{fig:optimalSplit}
\end{figure}

Next, to satisfy (II), we select the partition $s^*$ which maximizes the prediction metric,
\begin{equation}
  s^* = \argmax_{\substack{s_k,\\ M_D(s_k) < M^*_D}} M_Q(s_k). 
  \label{eq:optimalSplit}
\end{equation}
The optimal partition for the results shown above is highlighted in figure \ref{fig:optimalSplit}. 

\subsection{Comparison of $s^*$ with $M^*_Q$}
Finally we are able to assess the model's ability to predict the QoI. After identifying the optimal
partition we compare the model's performance, in this ``worst case'' scenario, against the threshold $M^*_Q$. 

If $s^*$ fails to satisfy the threshold then we conclude that the model is invalid, given the observations available,
and should not be used to make predictions for the QoI.
\begin{figure}[h]
  \begin{center}
    \includegraphics[angle=-90,width=3in]{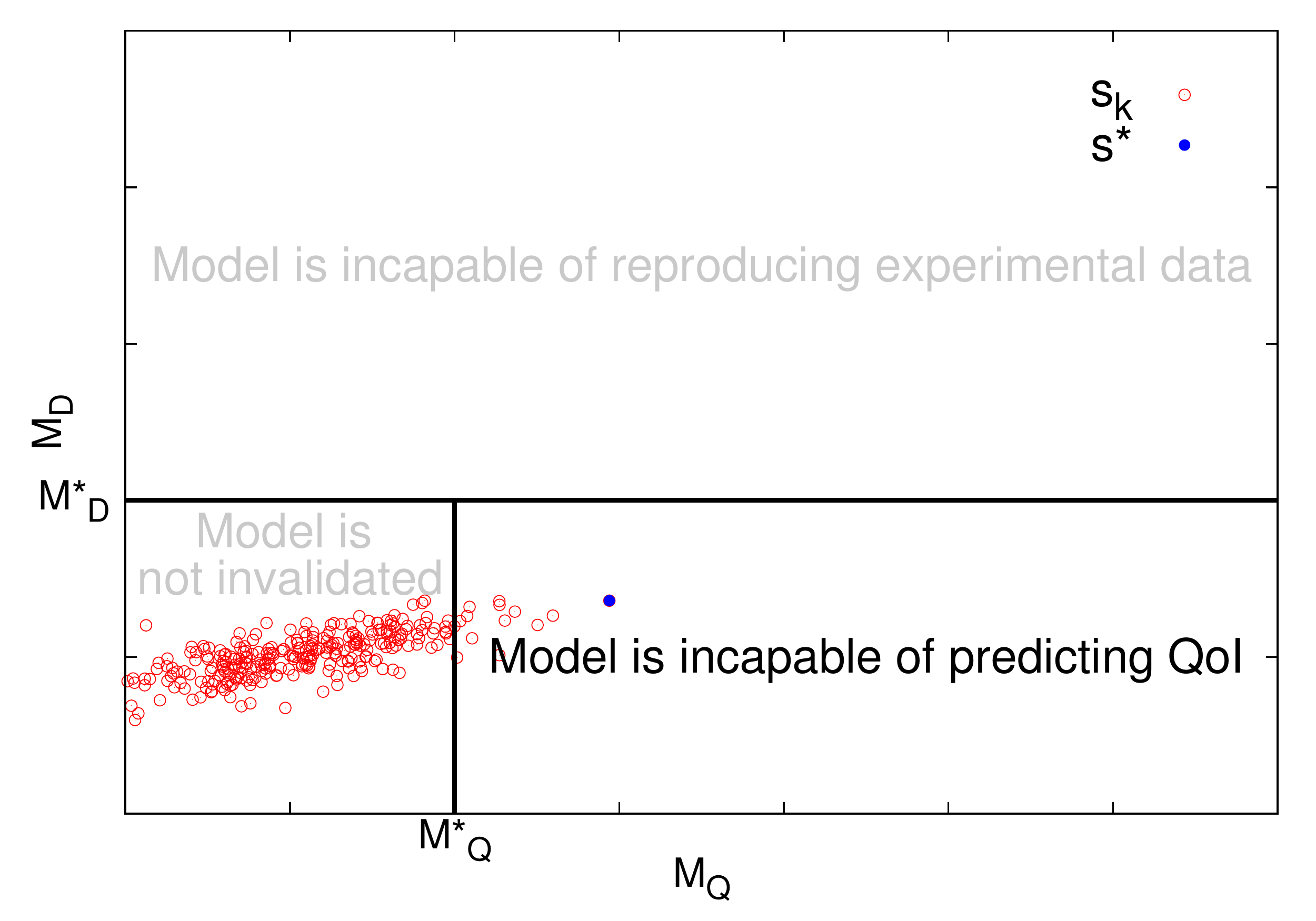}
  \end{center}
  \caption{Comparison of $s^*$ with $M^*_Q$}
  \label{fig:comparison}
\end{figure}
Figure \ref{fig:comparison} shows an example of exactly this case. 

If $s^*$ does not violate the tolerance set by the decision-maker we can only conclude that the model is not invalidated
given the observations we have. This does not guarantee that the model is valid, only that we cannot demonstrate
otherwise.  This outcome warrants further observations to continue challenging the model. The process to obtain these
additional experimental results may be supplemented by performing optimal experimental design \cite{Terejanu2011_ExpDes}. If, however, we cannot
obtain more data, then the process is complete, and we conclude that the model has not been invalidated.

\subsection{Algorithm}
The general algorithm can be summarized in 8 steps:
\begin{itemize}
  \item[1.] elicit the data metric and threshold, $M_D$ and $M_D^*$, from the modeler and/or experimentalist\\
  \item[2.] elicit the QoI metric and threshold, $M_Q$ and $M_Q^*$, from the decision-maker\\
  \item[3.] given $N$, choose the calibration set size, $N_C$, such that $$N_{C}+N_{V} = N$$
  \item[4.] generate all possible partitions of the data $\{s_k\}_{k=1}^P$, where $$P = {N \choose N_C} = \frac{N!}{N_C!N_V!}$$
  \item[5.] solve $P$ inverse problems for splits $\{s_k\}_{k=1}^P$
  \item[6.] for each partition $s_k$, compute $M_D(s_k)$ and $M_Q(s_k)$
  \item[7.] find optimal split $$s^* = \argmax_{\substack{s_k,\\ M_D(s_k) < M^*_D}} M_Q(s_k)$$
  \item[8.] compare $M_Q(s^*)$ with $M^*_Q$
\end{itemize}

Once more we stress that the procedure described above is extremely general. This is an advantage of the
approach and allows for its application to a wide range of problems. Specifics will be discussed for one application
below.


\section{Application to the data reduction model}

We now turn to a specific implementation of the proposed algorithm. A data reduction model is chosen for analysis.
Since higher level models require predictions obtained from this data reduction model, it is critical that we accurately
assess the quality of its predictions. Moreover, we must consider all partitions of the data because the model is
significantly influenced by the data. 

The inverse problem of calibrating the model parameters from the measurement data is solved
using Markov chain Monte Carlo simulations. In our simulations, samples from the posterior 
distribution are obtained using the
statistical library QUESO \cite{PrSc11} equipped with the Hybrid Gibbs
Transitional Markov Chain Monte Carlo method proposed by Cheung and Beck
\cite{Cheung_2009C}.

First, we briefly describe the experimental set-up, the model in question, and the quantity of interest. Next, we
describe the application of the algorithm and present the results.

\subsection{The ICCD camera}

\begin{figure}[h]
\begin{center}
\includegraphics[width=3in,height=0.9in]{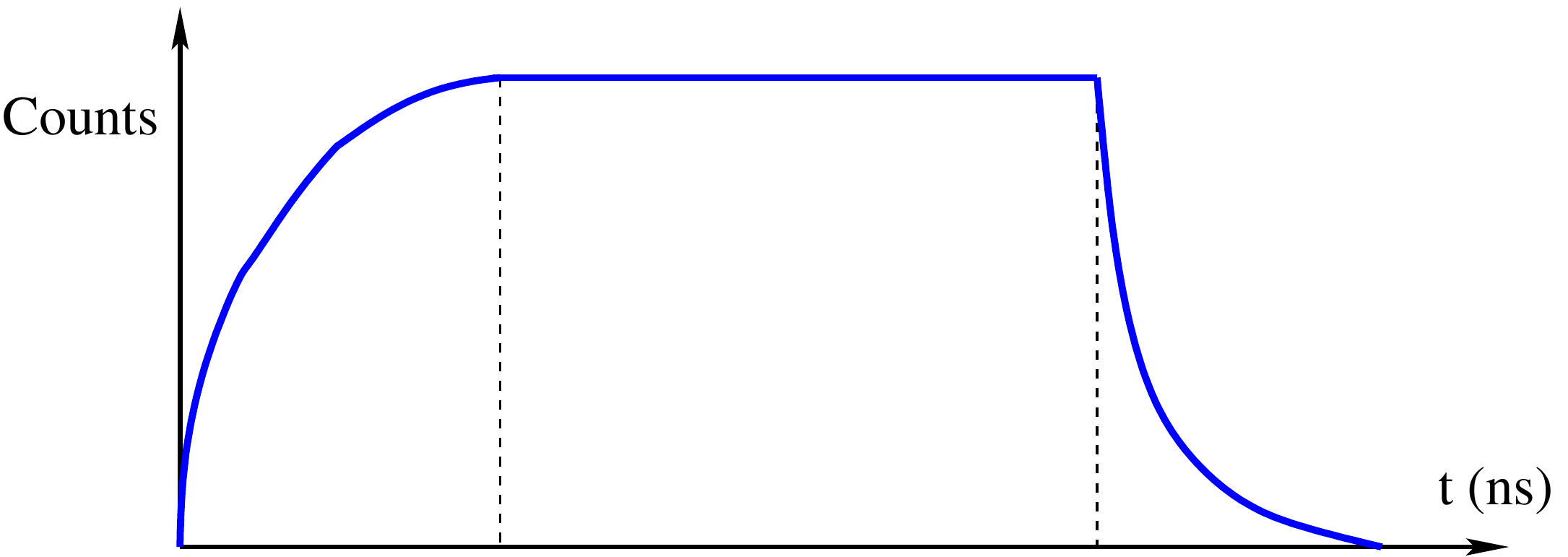}
\end{center}
\caption{Photon count vs. time for a small gate width in the ICCD camera}
\label{recip}
\end{figure}

The example problem comes from a shock-tube experiment in which an ICCD camera measures photon counts \cite{camera}. As the opening time, or gate width, of the camera increases, so does the photon count. For sufficiently large gate widths, this behavior is linear and simple to model. However, at very small gate widths, below the linear regime of the instrument, the behavior becomes complicated and nonlinear. Figure \ref{recip} shows a diagram of this behavior. It is in this nonlinear region where we wish to predict the photon count.

The raw data are photon counts received by the camera at eleven different gate widths ranging from
0.5 to 10 micro-seconds.  Given the photon count, $N_{\Delta t},$ computing the quantity of interest is a simple
post-processing step. The reciprocity $\rho$ is defined as
\begin{equation}
\rho(\Delta t)=\left(\frac{N_{10}}{10}\right) / \left(\frac{N_{\Delta t}}{\Delta t}\right),
\label{recipEqn}
\end{equation}
and the QoI is $\rho(0.1)$.
A data reduction model for the photon count as a function of gate width, $N_{\Delta t}$, is required because we do not have the data $N_{0.1}$.

If we only wished to predict the photon count in the linear regime of the instrument, we could simply use: 
\begin{equation}
N_{\Delta t} = \beta \Delta t.
\label{model1}
\end{equation}
 But in order to account for the nonlinearity encountered at very small gate widths, several new parameters are
 introduced into the model:
\begin{equation}
N_{\Delta t} = \beta (\Delta t + \delta) - \beta\frac{(\alpha_{1} - \alpha_{2})}{(\alpha_1 \alpha_2)}(1 - e^{-\alpha_{1}(\Delta t + \delta)}). 
\label{model3}
\end{equation}
Here, $\beta$ describes the linear term as before, and $\delta$ is a correction to the gate width $\Delta t$, in case it
is reported incorrectly. Also, $\alpha_{1}$ and $\alpha_{2}$ allow for differing opening and closing rates of the
camera, respectively. These four parameters must later be calibrated through the inverse problem.
\subsection{Application of the algorithm}
Now we apply the proposed algorithm to the example problem. The steps are as follows:

\begin{itemize}
\item[1.] As mentioned above, predictions of the quantity of interest result in cumulative distribution functions (CDFs)
in the units of the QoI. Thus, we compare the maximum horizontal distance between the CDFs, with some cutoff at the
tails \cite{Babuska2008}. In general, for two CDFs $F$ and
$G$, we define: 
\[ M_{Q} = \max_{u \in (\frac{\epsilon}{2}, 1- \frac{\epsilon}{2})} |x_{F,u} - x_{G,u}|\] 

\[ \text{where} \quad x_{F,u} = \min \{x \in \mathbb{R}; F(x) \geq u\}.\]

Here, $F$ is the posterior CDF using the calibration data, $G$ is the posterior CDF using all the data (including the validation data), and $\epsilon$ is a parameter between 0 and 1 controlling the amount of cut-off of the tails. See figure \ref{MQ}.
 
Using this metric, the tolerance for the quantity of interest is set. In this example, the threshold is $M_Q^* = 2$, which corresponds to the rather large percent error of about $100 \%.$

\begin{figure}[h]
\begin{center}
\includegraphics[width=3in]{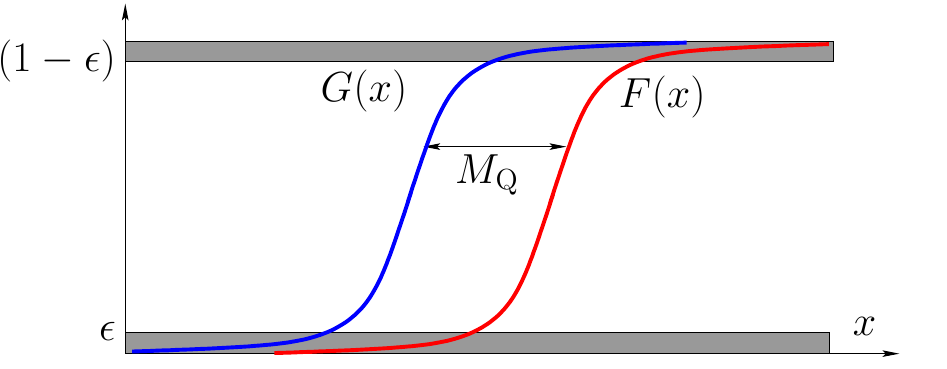}
\caption{the metric for the QoI}
\label{MQ}
\end{center}
\end{figure}

\item[2.]  The model's performance with respect to the data is characterized using a normalized difference between the
true observations and the corresponding predicted values: $$M_{D} = \mathbb{E} \left[ \sqrt{(X-D)^{T}\,\,
diag(D)^{-1}\,\, (X-D) }\right]$$ where $D$ is a vector of all the data, and $X$ is a vector of the predicted values
corresponding point-wise to $D$. Note that the entries of $X$ are the predicted values of the observables using the
calibrated model (on just the calibration set), but for \textit{all} values of gate widths (including those
corresponding to both calibration and validation sets).
 
Using the above as the data metric, the tolerance $M^*_D$ is set to 0.2, which yields an average relative error of
20\%.

\item[3.]  Given 11 gate widths, we choose calibration set size $N_{C}=7$, leaving the validation set size as $N_{V}=4$. In fact, as described earlier, multiple data points are provided for each of the gate widths. However, these points are not considered individually during the partioning process, but are kept together as a single unit.

\item[4.]  We generate all possible data sets for calibration: \[ {11\choose 7} = 330.\]

\item[5.]  We solve these 330 inverse problems. Again, this is done probabilistically using uniform priors on all parameters and Bayesian updating.

\item[6.]  We compute $M_{Q}$ and $M_{D}$ for each of the 330 splits of the data. Results are shown in figure
\ref{dataplot}.

\item[7.]  We find all points which satisfy $M_{D}(s_k) < M^{*}_{D}$. In this example, all points satisfy this
requirement. Next, we find \[s^* = \argmax_{\substack{s_k,\\ M_D(s_k) < M^*_D}} M_Q(s_k).\] This is the right-most point
on the plot, shown in blue
in figure \ref{dataplot}.

\begin{figure}[h]
\begin{center}
\includegraphics[angle=-90,width=3in]{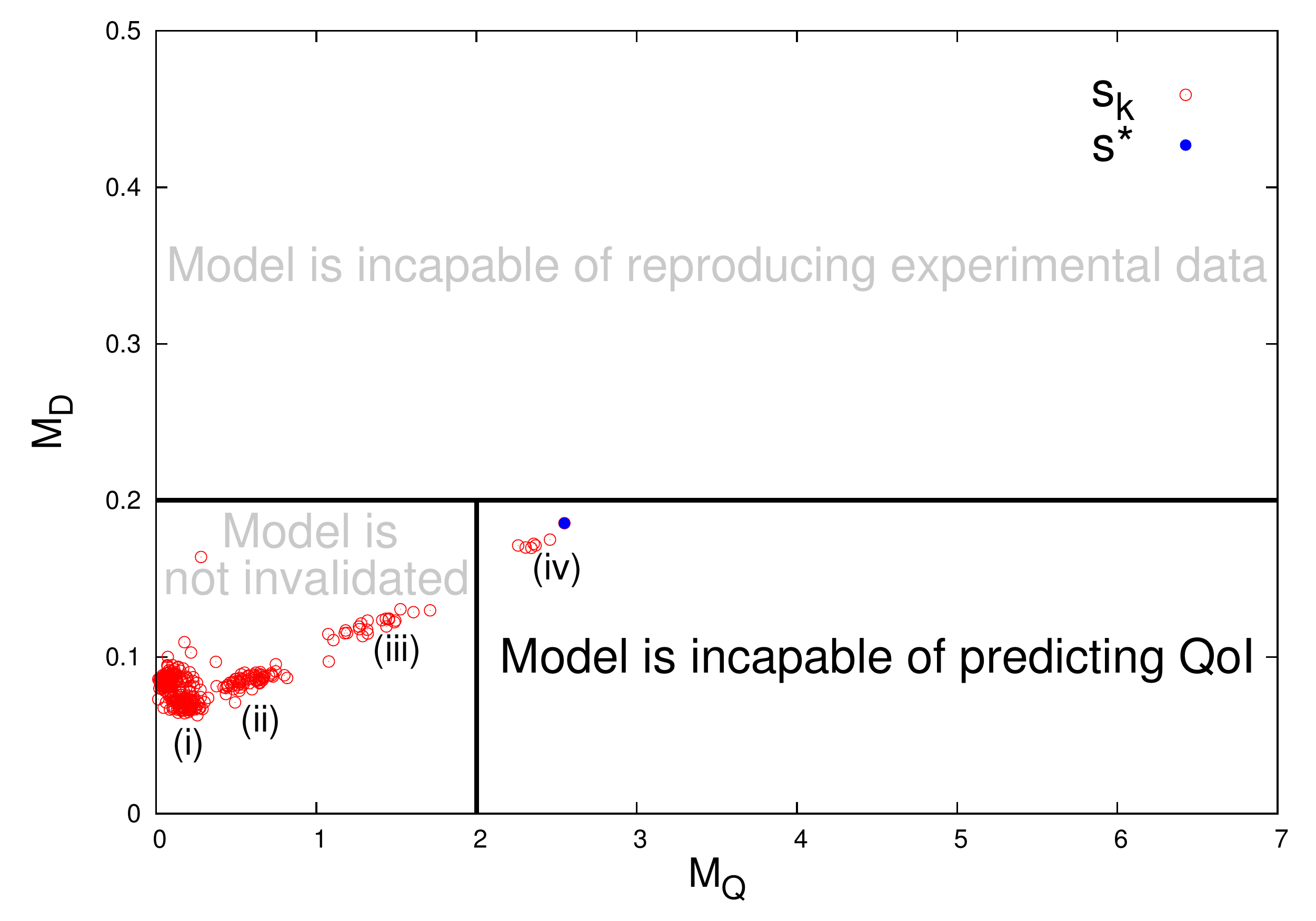}
\end{center}
\caption{Results as described in step 7.}
\label{dataplot}
\end{figure}
\item[8.] We compare $M_Q(s^*)$ with $M^{*}_{Q}$. Since $M_Q(s^*) > M^{*}_{Q}$, then we must conclude that the model is incapable of predicting the QoI, and has thus been invalidated.
\end{itemize}

\subsection{Analysis of results}
Looking at figure \ref{dataplot}, we see that the points, representing splits, are grouped into four regions.  First, we
note that $s^*$ belongs to group (iv), and as group number increases, the size of the group decreases.  Now let us
examine why $s^*$ is the most challenging of the splits and why these groups have formed.

First, recall that the data we received was $N_{\Delta t}$, where $\Delta t$ ranged from 0.5 to 10 micro-seconds;
specifically $\Delta t =
0.5, 0.6, 0.7, 0.8, 0.9, 1, 2, 3, 4, 5,$ and $10\,\, \mu$s. As one might guess, the most challenging
(optimal) data split, $s^*$, contained the four lowest gate widths in the validation set.
In other words, the data points ``closest'' to the QoI scenario were 
missing from the calibration set. After calibration, the model must first predict the QoI without
any of the information contained in the lowest gate widths. Then, using the
validation points to recalibrate the model, there is a large distance between the resulting predictions (CDFs) for the
QoI. 

The calibration set for $s^*$ is missing $0.5, 0.6, 0.7, 0.8$. The other points in group (iv) are those whose
calibration set is missing $0.5, 0.6, 0.7$, but at least contains $0.8$, making them slightly less challenging than $s^*$.
Next, group (iii) contains those splits whose calibration set is missing $0.5$ and $0.6$. Similarly, in group (ii), we
see all the splits whose calibration set is missing $0.5$. Finally, in group (i), the calibration sets contain $0.5$,
and some collection of the remaining points. 

The grouping of the data suggests a possible way to decrease the number of inverse problems performed. By finding
representatives of the groups, we could search for the optimal split without analyzing every one. Moreover, in the case
that the model is not invalidated, understanding these groups could be helpful when planning further experiments,
perhaps through experimental design.


\section{Conclusion}
Computationally, our approach is prohibitively expensive and requires significant improvement. Methods to reduce the
number of inverse problems required are being investigated. Among them is the use of mutual information to group
observations into sets containing similar data. These larger sets then require fewer inverse problems. 

With experimental design one may be able to choose where to perform subsequent experiments that will challenge the model
further, finding a new optimal split. This newly determined split could render the model invalid, or provide increased
confidence in the model's predictive capability. 

This paper has proposed and demonstrated a systematic framework for assessing a model's predictive performance with
respect to a particular quantity of interest. Extending the work of Babu\v{s}ka et al., the ability of the model to
reproduce experimental observations was also evaluated. 

A data reduction model was examined using our framework and ultimately deemed invalid. While the model was capable of
reproducing the observations at higher gate widths, it failed based on its performance in predicting the quantity of
interest.  The analysis was carried out on all partitions of the data respecting a chosen size constraint. This allowed
for the determination of an optimal set satisfying calibration (I) and validation (II) requirements.



%
\nocite{*}


\bibliographystyle{BibTeXtran}   
\bibliography{paper}       

\end{document}